\documentclass[english]{article}
\usepackage[T1]{fontenc}
\usepackage[latin9]{inputenc}
\usepackage{array}
\usepackage{float}
\usepackage{amsmath}
\usepackage{amssymb}

\makeatletter

\providecommand{\tabularnewline}{\\}

\usepackage{amsfonts,amssymb}
\usepackage{latexsym}
\usepackage{epsfig}
\usepackage{cite}
\usepackage{graphicx}
\usepackage[colorlinks,linkcolor=blue]{hyperref}
\newcommand{\be}{\begin{equation}}
\newcommand{\ee}{\end{equation}}

\makeatother

\usepackage{babel}
\begin{document}
{}~ \hfill\vbox{\hbox{CTP-SCU/2020010}}\break
\vskip 3.0cm
\centerline{\Large \bf  Thermodynamics and Weak Cosmic Censorship Conjecture }
\vspace*{1.0ex}
\centerline{\Large \bf  of 4D Gauss-Bonnet-Maxwell Black Holes}
\vspace*{1.0ex}
\centerline{\Large \bf via Charged Particle Absorption}

\vspace*{10.0ex}
\centerline{\large Shuxuan Ying}

\vspace*{4.0ex}
\centerline{\large \it Department of Physics, Chongqing University,}
\centerline{\large \it Chongqing 401331, China}\vspace*{1.0ex}
\vspace*{2.0ex}
\centerline{\large \it Center for Theoretical Physics, College of Physics,}
\centerline{\large \it Sichuan University, Chengdu, 610064, PR China}\vspace*{1.0ex}
\vspace*{4.0ex}

\centerline{ysxuan@stu.scu.edu.cn}
\vspace*{10.0ex}
\centerline{\bf Abstract} \bigskip \smallskip
Recently, the non-trivial solutions for 4-dimensional black holes of Einstein-Gauss-Bonnet gravity had been discovered. In this paper, considering a charged particle entering into a 4-dimensional Gauss-Bonnet-Maxwell black hole, we calculate the black hole thermodynamic properties by using the Hamilton-Jacobi equation. In the normal phase space, the cosmological constant and Gauss-Bonnet parameter are fixed, the black hole satisfies the first and second laws of thermodynamics and the weak cosmic censorship conjecture (WCCC) is valid. On the other hand, in the case of extended phase space, the cosmological constant and Gauss-Bonnet parameter are treated as the thermodynamic variables. The black hole also satisfies the first law of thermodynamics. However, the increase or decrease of  black hole's entropy depends on some specific conditions. Finally, we observe that the WCCC is violated for the near-extremal black holes in the extended phase space.
\vfill
\eject
\baselineskip=16pt
\vspace*{10.0ex}
\tableofcontents

\section{Introduction}

Studying the thermodynamic properties of black holes helps us understand
the quantum gravity. Considering a particle with a negative energy
entering into spacetime with an ergoregion (e.g. a rotating black hole), the energy of the black
hole would be extracted \cite{Penrose:1969pc}. Later, Christodoulou
found that there was an irreducible mass when a black hole absorbed
a particle \cite{Christodoulou:1970wf,Bardeen:1970zz,Christodoulou:1972kt}.
Due to a relationship between the irreducible mass and the entropy,
the entropy of the black hole corresponds to its area of horizon (or
the square of its irreducible mass) \cite{Bekenstein:1972tm,Bekenstein:1973ur}.
Furthermore, Hawking et al. established the four laws of thermodynamics
for black holes which are similar to those in the statistical thermodynamics
\cite{Bardeen:1973gs}. Based on the contributions above, the thermodynamic
properties of black holes can be calculated semi-classically without
considering the complete quantum gravity. In addition, Hawking found
that the black holes have the temperatures in curved spacetime \cite{Hawking:1974sw}.

When Maldacena discovered the AdS/CFT correspondence \cite{Maldacena:1997re},
the studies on the asymptotic anti-de Sitter (AdS) black holes became
very popular. Hawking and Page presented a first-order phase transition
between the thermal AdS space and the Schwarzschild AdS black hole
\cite{Hawking:1982dh}. In the following work \cite{Witten:1998zw},
this result was translated to the language of AdS/CFT, which is called
a confinement/deconfinement phase transition. More results on the
thermodynamics and phase structures of AdS black holes can be found
in refs. \cite{Chamblin:1999tk,Chamblin:1999hg,Caldarelli:1999xj,Cai:2001dz,Kubiznak:2012wp}.
Recently, when considering the asymptotical AdS black holes in the
the extended phase space, an interesting idea has been proposed to
explain the cosmological constant in terms of a thermodynamic pressure
\cite{Kastor:2009wy,Dolan:2011xt}. Next, many thermodynamic studies
on various AdS black holes in the the extended phase space were presented \cite{Gunasekaran:2012dq,Wei:2012ui,Cai:2013qga,Xu:2014kwa,Frassino:2014pha,Hennigar:2015esa,Wang:2018xdz}
.

Moreover, the weak cosmic censorship conjecture (WCCC) states that
the singularity is always in the event horizon \cite{Penrose:1969pc}.
If an observer is from the normal initial conditions, then the singularity
cannot be observed in the future infinity in any real physical process.
Wald proposed a method that threw a test particle into a black hole
in order to test the validity of the WCCC in an extremal Kerr-Newman
black hole \cite{Wald}. However, unfortunately, due to the electromagnetic
or centrifugal repulsion, the black hole cannot capture the test particle
with the sufficiently large charge or angular momentum to overcharge
or overspin itself. Moreover, considered the near-extremal charged/rotating
black hole, overcharged/overspin black hole by absorbing a particle
could be observed \cite{Hubeny:1998ga,Jacobson:2009kt,Saa:2011wq}.
However, because of the backreaction and self-force, the WCCC still
be true for these black holes \cite{Hod:2008zza,Barausse:2010ka,Barausse:2011vx,Zimmerman:2012zu,Colleoni:2015afa,Colleoni:2015ena}.
This is worth noting that there is no strong evidence to prove the
WCCC, many papers discussed its validity in various black holes \cite{Matsas:2007bj,Richartz:2008xm,Isoyama:2011ea,Gao:2012ca,Hod:2013vj,Duztas:2013wua,Siahaan:2015ljs,Natario:2016bay,Duztas:2016xfg,Revelar:2017sem,Sorce:2017dst,Husain:2017cmj,Crisford:2017gsb,Gwak:2017kkt,An:2017phb,Ge:2017vun,Yu:2018eqq,Gwak:2018akg,Gim:2018axz,Chen:2018yah,Zeng:2019jta,Chen:2019nsr,Gwak:2019asi,Zeng:2019jrh,Chen:2019pdj,Wang:2019jzz,Zeng:2019hux,Hong:2019yiz,Hu:2019rpw,Hu:2019qzw,Mu:2019bim,He:2019fti,Gan:2019jac,Zeng:2019baw,He:2019kws,Hu:2019zxr,Gwak:2019rcz,Gan:2019ibg,Hong:2020zcf,Gwak:2020zht,Wang:2020osg}.
In particular, the Reissner-Nordstrom (RN)-AdS black hole via the
charged particle absorption was considered in the normal and extended
phase spaces \cite{Zhang:2013tba,Gwak:2017kkt}. The first law of
thermodynamics and the WCCC were satisfied, meanwhile the second law
of thermodynamics holds near the extreme value. It is noteworthy that
the second law of thermodynamics is always valid for a RN-AdS black
hole in the normal phase space. In the ref. \cite{Zeng:2019hux},
considering a charged particle absorption, the authors studied the
thermodynamics and the WCCC of Gauss-Bonnet AdS Black Holes in higher
dimensions ($D>4$).

On the other hand, it is well known that the non-trivial static spherically
symmetric solutions only exist in the Gauss-Bonnet gravity when the
number of spacetime dimension $D>4$. Otherwise, when $D=4$, the
Gauss-Bonnet term becomes a topological invariant which does not contribute
to the gravitational equations of motion. In recent work \cite{Glavan:2019inb},
the modified Einstein-Gauss-Bonnet gravity was proposed for obtaining
the non-trivial solutions in the limit $D\rightarrow4$. The method
was to mimic the dimensional regularization in quantum field theory
and rescaled the coupling parameter $\alpha$ to $\alpha/\left(D-4\right)$.
As a result, the divergence of the Gauss-Bonnet contribution when
$D=4$ could be canceled by the factor $\left(D-4\right)$. And then
the non-trivial four dimensional black hole solutions with Gauss-Bonnet
contribution were obtained. These new black hole solutions open a
new window to study the Gauss-Bonnet effect in the lower dimensional
theory.
It is worth noting that although the divergence of the variation of the Gauss-Bonnet action had been cancelled out and introduced a brand-new black hole solution, its thermodynamic properties still have some problems in the limit $D\rightarrow4$. Therefore, the coupling replacement $\alpha\rightarrow\alpha/\left(D-4\right)$ cannot describe the topologically nontrivial solutions, and then its thermodynamic result is preliminary and requires the future's lots of studies \cite{Lu:2020iav}.
On the other hand, traditionally, constructing the 4-dimensional dynamics by dimensional
reduction from higher-dimensional theory, the degrees of freedom of
the extra dimensions cannot be discarded. And the additional fields
must be considered in the 4-dimensional theory \cite{Lu:2020iav}.
Furthermore, in ref.  \cite{Ai:2020peo}, to understand how dimensional regularization throws away the dynamics of the extra dimensions,
the author embedded the 2-dimensional spacetime
into D-dimensional spacetime to figure out the map between the metric
tensors and the Einstein equations in 2-dimensional and higher dimensional theories.
And then considering the limit, the dynamical equations from extra
dimensions could be discarded. This limit procedure requires highly
symmetric backgrounds which lead the embedment to be valid. However,
it is still unclear how D-dimensional dynamical equations converges
to the 4-dimensional dynamical equations by taking the limit $D\rightarrow4$.
Moreover, it is difficult to figure
out which dimensions retain in the final action. Because the limit
$D\rightarrow4$ is only achieved by replacing the coupling constant
$\alpha\rightarrow\alpha/\left(D-4\right)$ but not modifying the
Riemann tensors. This replacement cancels out the vanishing factor
of the Gauss-Bonnet term, and therefore introduce the local
dynamics from the Gauss-Bonnet contribution.
In addition, the solutions of the
Gauss-Bonnet gravity with Maxwell theory in AdS space were given and
some thermodynamic properties were discussed in ref. \cite{Fernandes:2020rpa}.
Then, a lot of work on the thermodynamic properties of 4-dimensional
Gauss-Bonnet black holes appeared. In ref. \cite{Hegde:2020xlv},
the thermodynamics and the phase structure in the extended phase space
was discussed, where the cosmological constant was seen as the thermodynamic
pressure. And the critical behaviors were studied in refs. \cite{Singh:2020xju,HosseiniMansoori:2020yfj}.
Furthermore, the authors studied the phase transition and microstructures
for the four-dimensional charged AdS black hole in ref. \cite{Wei:2020poh}.
In ref. \cite{EslamPanah:2020hoj}, 4-dimensional Einstein-Gauss-Bonnet
AdS black holes were treated as heat engines. In addition to thermodynamic
properties, the stabilities of 4-dimensional Gauss-Bonnet black holes
were also discussed in recent works \cite{Konoplya:2020bxa,Konoplya:2020juj,Li:2020tlo,Zhang:2020sjh}.
And in refs. \cite{Konoplya:2020bxa,Mishra:2020gce} , quasinormal
modes and strong cosmic censorship are considered.

In this paper, based on the developments above, it ought to be studying
the thermodynamics and the WCCC for this new 4-dimensional Gauss-Bonnet-Maxwell
black hole. To be specific, we are trying to verify the thermodynamic laws and the WCCC by throwing a test particle into to an over-charged 4-dimensional Gauss-Bonnet-Maxwell black hole not only in the normal phase space but also in the extended phase space. In the extended phase space, the cosmological constant and Gauss-Bonnet parameter are both treated as thermodynamic variables, which leads to some interesting results. For example, in the extended space, the black hole thermodynamics in Gauss-Bonnet gravity with a quadratic nonlinear electrodynamics are discussed in ref. \cite{Hendi:2015oqa} and the thermodynamic properties of Gauss-Bonnet-Born-Infeld-massive black holes are studied in ref. \cite{Hendi:2016yof}. On the other hand, in ref. \cite{EslamPanah:2017yoc}, the authors considered the thermodynamic behavior of the Gauss-Bonnet-massive gravity with the power-Maxwell field in the normal space. In this paper, in the normal phase space, when a charged particle enters into a 4-dimensional Gauss-Bonnet-Maxwell black hole, the first and second laws of thermodynamics are still satisfied. However, the results are different in the extended phase space. Considering the thermodynamic laws in the extended phase space, the first law of thermodynamics is also satisfied, but the second law is indefinite. If we assume the Gauss-Bonnet parameter dose not change after the black hole absorbs a charged particle, the second law is violated for the extremal and near-extremal black holes. On the other hand, considering the WCCC, there are still some different in the normal and extended phase space. For near-extremal black holes, the WCCC is still valid in the normal phase space. However for the extended phase space, the WCCC is violated. Furthermore, after absorbing the charged particle, the extremal black hole becomes non-extremal in the normal phase space, but the extremal black hole stays extremal in the extended phase space.

The rest of this paper is organized as follows. In section
\ref{Sec:GB-NLED-BH}, we derive the general solutions of the 4-dimensional
Gauss-Bonnet-Maxwell black holes, and its thermodynamic properties.
In section \ref{Sec:TCPA}, we first review the Hamilton-Jacobi equation
for a particle entering into an black hole in curved spacetime. Then
we obtain the thermodynamics of the black hole in the normal and extended
phase space, and discuss the first law and the second law of thermodynamics.
In section \ref{Sec:WCCC}, we throw a charged particle into the black
hole to test the WCCC in the both phase spaces. In section \ref{Sec:Con},
we give the conclusions. We take $G=\hbar=c=k_{B}=1$ for simplicity
in this paper.

\section{4-dimensional Gauss-Bonnet-Maxwell gravity}

\label{Sec:GB-NLED-BH}

In this section, we briefly review the Gauss--Bonnet gravity coupled
to the Maxwell theory in 4-dimensions. And we will also discuss its
thermodynamic properties. Based on ref.\cite{Glavan:2019inb}, the
Gauss-Bonnet-Maxwell theory is described by the action

\begin{equation}
\mathcal{S}=\frac{1}{16\pi}\underset{\mathcal{M}}{\int}d^{D}x\sqrt{-g}\left[R-2\Lambda+\frac{\alpha}{D-4}\left(R^{2}-4R_{\mu\nu}R^{\mu\nu}+R_{\mu\nu\rho\sigma}R^{\mu\nu\rho\sigma}\right)-F^{\mu\nu}F_{\mu\nu}\right],\label{eq:Action}
\end{equation}
where $\Lambda=-\frac{\left(D-1\right)\left(D-2\right)}{2l^{2}}$
is the cosmological constant, $l$ is the AdS radius, $\alpha$ is
the Gauss-Bonnet coupling constant, and $F_{\mu\nu}=\partial_{\mu}A_{\nu}-\partial_{\nu}A_{\mu}$
is the electromagnetic field-strength tensor in which $A_{\mu}$ is
the gauge potential. Varying the action $\eqref{eq:Action}$ yields
the equations of motion,
\begin{eqnarray}
R_{\mu\nu}-\frac{1}{2}\left(R-2\varLambda\right)g_{\mu\nu}+H_{\mu\nu} & = & 8\pi T_{\mu\nu},\nonumber \\
\nabla_{\mu}F^{\mu\nu} & = & 0,
\end{eqnarray}
where
\begin{eqnarray}
H_{\mu\nu} & = & -\frac{1}{2}\frac{\alpha}{D-4}\left(R^{2}-4R_{\mu\nu}R^{\mu\nu}+R_{\mu\nu\rho\sigma}R^{\mu\nu\rho\sigma}\right)g_{\mu\nu}\nonumber \\
 &  & +2\frac{\alpha}{D-4}\left(RR_{\mu\nu}-2R_{\mu\alpha}R^{\alpha\beta}g_{\beta\nu}-2R_{\mu\lambda\nu\sigma}R^{\lambda\sigma}+g_{\beta\nu}R_{\mu\gamma\sigma\lambda}R^{\beta\gamma\sigma\lambda}\right),\\
T_{\mu\nu} & = & \frac{1}{4\pi}\left(-\frac{1}{4}F^{\mu\nu}F_{\mu\nu}g_{\mu\nu}+F_{\mu}^{\,\lambda}F_{\nu\lambda}\right).\nonumber
\end{eqnarray}
Considering a 4-dimensional static spherically symmetric black hole
ansatz, we take the following metric and vector potential
\begin{eqnarray}
ds^{2} & = & -f\left(r\right)dt^{2}+\frac{dr^{2}}{f\left(r\right)}+r^{2}\left(d\theta^{2}+\sin^{2}\theta d\phi^{2}\right)\text{,}\nonumber \\
A & = & A_{t}\left(r\right)dt\text{.}\label{eq:ansatz}
\end{eqnarray}
After setting $D\rightarrow4$, the equations of metric function $f\left(r\right)$
and vector potential $A_{t}\left(r\right)$ are written as
\begin{eqnarray}
0 & = & 1-f\left(r\right)-rf^{\prime}\left(r\right)+\frac{3}{l^{2}}r^{2}++2\alpha f^{\prime}\left(r\right)\left(f\left(r\right)-1\right)r^{-1}\nonumber \\
 &  & -\alpha\left(f\left(r\right)-1\right)^{2}r^{-2}+\left(\partial_{r}A_{t}\left(r\right)\right)^{2}r^{2},\label{eq:tt}\\
0 & = & \left[r^{2}\partial_{r}A_{t}\left(r\right)\right]^{\prime}.\label{eq:Maxwell}
\end{eqnarray}
By solving eqn. (\ref{eq:tt}) and eqn. (\ref{eq:Maxwell}), one can
obtain the solutions for metric function and vector potential \cite{Fernandes:2020rpa}
\begin{eqnarray}
f\left(r\right) & = & 1+\frac{r^{2}}{2\alpha}\left[1-\sqrt{1+4\alpha\left(-\frac{1}{l^{2}}+\frac{2M}{r^{3}}-\frac{Q^{2}}{r^{4}}\right)}\right],\label{eq:f(r)}\\
A_{t}\left(r\right) & = & -\frac{Q}{r},\label{eq:vector potential}
\end{eqnarray}
where $M$ is the ADM mass of the black hole and $Q$ is the black
hole charge. The thermodynamic properties of the black hole can be
defined on the black hole horizon $r_{+}$ , which is determined by
$f\left(r_{+}\right)=0$. The Hawking temperature, electrical potential
and entropy of the black hole are given by \cite{Fernandes:2020rpa,Wei:2020poh}
\begin{eqnarray}
T & \equiv & \frac{f^{\prime}\left(r_{+}\right)}{4\pi}=\frac{-\alpha+r_{+}^{2}+3\frac{r_{+}^{4}}{l^{2}}-Q^{2}}{4\pi\left(r_{+}^{3}+2\alpha r_{+}\right)},\label{eq:HT}\\
\Phi & \equiv & \int_{r_{+}}^{\infty}A_{t}^{\prime}\left(r\right)=-A_{t}\left(r_{+}\right),\label{eq:potential}\\
S & \equiv & \int\frac{dM}{T}=\pi r_{+}^{2}+4\pi\alpha\ln\frac{r_{+}}{\sqrt{\alpha}},\label{eq:entropy}
\end{eqnarray}
where $A_{t}\left(r\right)$ goes to zero at $r=\infty$, the electrostatic
potential $\Phi$ is a conjugated thermodynamic variable to black
hole charge $Q$. It is worth noting that $\sqrt{\alpha}$ in eqn.
(\ref{eq:entropy}) is from the identification with an integral constant,
and the purpose of this identification is to ensure $\ln\frac{r_{+}}{\sqrt{\alpha}}$
for dimensionless and the Smarr relation (\ref{eq:Smarr relation})
associated with the entropy is consistent with the higher-dimensional
form.

\section{Thermodynamics via Charged Particle Absorption}

\label{Sec:TCPA}

In this section, we study the black hole thermodynamics of the Einstein-Gauss-Bonnet
gravity coupled to the Maxwell theory through a charged particle entering
the black hole horizon. Because of the conservations of energy and
charge, after absorbing a charged particle, the mass and charge of
the black hole would change too. Furthermore, the other thermodynamic
quantities of the black hole may also change. The purpose of this
section is to check whether the change in the thermodynamic variables
of the black hole will violate the first and second laws of thermodynamics
in the normal and extended phase spaces.

At first, we briefly review the relationship of the test particle's
energy with its radial momentum and potential energy before the particle
enters the horizon. The Hamilton-Jacobi equation of the test particle
is given by \cite{Wang:2019jzz}:
\begin{equation}
-\frac{\left[E+qA_{t}\left(r\right)\right]^{2}}{f\left(r\right)}+\frac{\left[P^{r}\left(r\right)\right]^{2}}{f\left(r\right)}+\frac{L^{2}}{r^{2}}=m^{2}\text{,}\label{eq:HJE}
\end{equation}
where $L$ is the particle's angular momentum and $P^{r}\left(r\right)$
is the particle's radial momentum. It is worth mentioning that $P^{r}\left(r_{+}\right)$
is finite and proportional to the Hawking temperature of the black
hole \cite{Benrong:2014woa,Tao:2017mpe}. Since the energy of the
particle is required to be a positive value \cite{Christodoulou:1970wf,Christodoulou:1972kt},
we can rewrite eqn. (\ref{eq:HJE}) as
\begin{equation}
E=-qA_{t}\left(r\right)+\sqrt{f\left(r\right)\left(m^{2}+\frac{L^{2}}{r^{2}}\right)+\left[P^{r}\left(r\right)\right]^{2}}.
\end{equation}
At the horizon $r=r_{+}$, the above equation reduces to
\begin{equation}
E=q\Phi+\left\vert P^{r}\left(r_{+}\right)\right\vert ,\label{eq:Ehorizon}
\end{equation}
which relates the energy of the particle to its radial momentum and
potential energy just before the particle enters the horizon.

For convenience, before the subsequent discussions on the thermodynamic
properties via charged particle absorption, we write down the following
formulas:

\begin{eqnarray}
\frac{\partial f\left(r\right)}{\partial r}|_{r=r_{+}} & = & 4\pi T\text{, }\nonumber \\
\frac{\partial f\left(r\right)}{\partial M}|_{r=r_{+}} & = & -\frac{2}{r_{+}+\frac{2\alpha}{r_{+}}}\text{,}\nonumber \\
\frac{\partial f\left(r\right)}{\partial l}|_{r=r+} & = & -\frac{2r_{+}^{2}}{l^{3}}\frac{1}{1+\frac{2\alpha}{r_{+}^{2}}}\text{,}\label{eq:fdrh}\\
\frac{\partial f\left(r\right)}{\partial Q}|_{r=r_{+}} & = & \frac{2\Phi}{r_{+}+\frac{2\alpha}{r_{+}}}\text{,}\nonumber \\
\frac{\partial f\left(r\right)}{\partial\alpha}|_{r=r_{+}} & = & \frac{1}{r_{+}^{2}+2\alpha}.\nonumber
\end{eqnarray}

\subsection{Normal Phase Space}

Considering the normal phase space, only the black hole mass $M$
and black hole charge $Q$ are the thermodynamic variables. We assume
that the charged particle, which enters into the horizon of the black
hole, have the energy $E$ and charge $q$. And the black hole changes
from $\left(M,Q\right)$ to $\left(M+dM,Q+dQ\right)$ after absorbing
a charged particle. In the normal phase space, the black hole mass
$M$ is considered as the internal energy $U$ of the black hole.
Based on the law of conservation of energy and charge, we have the
formulas:
\begin{equation}
dM=E\text{ and }dQ=q.\label{eq:dMdQNS}
\end{equation}
Before the charge particle enters the black hole horizon, the outer
horizon radius $r_{+}$ satisfies
\begin{equation}
f\left(r_{+};M,Q\right)=0.
\end{equation}
Then after absorbing the charge particle, the black hole horizon radius
is written as $r_{+}+dr_{+}$, which also satisfies
\begin{equation}
f\left(r_{+}+dr_{+};M+dM,Q+dQ\right)=0\text{.}
\end{equation}
Therefore, the total differential of $f$ can be obtained by
\begin{equation}
\frac{\partial f\left(r\right)}{\partial r}|_{r=r_{+}}dr_{+}+\frac{\partial f\left(r\right)}{\partial M}|_{r=r_{+}}dM+\frac{\partial f\left(r\right)}{\partial Q}|_{r=r_{+}}dQ=0\text{.}\label{eq:NPFL}
\end{equation}
Substituting eqns. (\ref{eq:fdrh}) into eqn. (\ref{eq:NPFL}), and
then combining eqn. (\ref{eq:entropy}) to remove the $dr_{+}$ term,
we can obtain the first law of thermodynamics
\begin{equation}
dM=\Phi dQ+TdS.\label{eq:1stNS}
\end{equation}
Furthermore, using eqn.(\ref{eq:Ehorizon}), (\ref{eq:dMdQNS}) and
(\ref{eq:1stNS}), the variation of entropy becomes
\begin{equation}
dS=\frac{\left\vert P^{r}\left(r_{+}\right)\right\vert }{T}>0\text{,}
\end{equation}
which means that absorbing a charged particle, in normal phase space,
does not violate the second law of thermodynamics.

\subsection{Extended Phase Space}

In the extended phase space, not only the black hole mass $M$ and
black hole charge $Q$, but also the cosmological constant $l$ and
the Gauss--Bonnet parameter $\alpha$ in metric function $f\left(r\right)$
are all the thermodynamic variables. In this case, we define the thermodynamic
pressure of the black hole by using the cosmological constant \cite{Kastor:2009wy,Dolan:2011xt}:

\begin{equation}
P\equiv-\frac{\Lambda}{8\pi}=\frac{3}{8\pi l^{2}}.\label{eq:P}
\end{equation}
And the conjugate thermodynamic volume of the black hole is given
by:
\begin{equation}
V=\left(\frac{\partial M}{\partial P}\right)_{S,Q,\alpha}=\frac{4\pi}{3}r_{+}^{3},\label{eq:V}
\end{equation}
where we use the eqn. $\eqref{eq:f(r)}$ and $f\left(r_{+}\right)=0$.
Furthermore, the conjugate quantity of Gauss-Bonnet parameter $\alpha$
is $\mathcal{A}$, which is defined as \cite{Wei:2020poh}:
\begin{equation}
\mathcal{A}=\left(\frac{\partial M}{\partial\alpha}\right)_{S,Q,P}=\frac{1}{2r_{+}}+2\pi T\left(1-2\ln\frac{r_{+}}{\sqrt{\alpha}}\right).\label{eq:conjugate of alpha}
\end{equation}
The Smarr formula hence can be confirmed as
\begin{equation}
M=2TS+\Phi Q-2PV+2\mathcal{A}\alpha,\label{eq:Smarr relation}
\end{equation}
which is consistent with the higher-dimensional form for Smarr formula
\cite{Cai:2013qga}. Moreover, in the extended phase space, the black
hole mass $M$ should be treated as the enthalpy $H$ instead of internal
energy $U$ of the black hole \cite{Cai:2013qga}. Therefore, in the
extended phase space, a charged particle of the energy $E$ and charge
$q$ enter the black hole horizon, which causes the internal energy
and charge of the black hole to change as
\begin{equation}
dU=d\left(M-PV\right)=E\text{ and }dQ=q.\label{eq:EQES}
\end{equation}
No matter the radius $r$ takes the initial black hole horizon radius
$r_{+}$ or the changed horizon radius $r_{+}+dr_{+}$ after absorbing
a charged particle, the metric function $f\left(r\right)$ must set
to be zero. Moreover, we can obtain the infinitesimal changes in $M$,
$Q$, $l$, $\alpha$ and $r_{+}$ :
\begin{equation}
\frac{\partial f\left(r\right)}{\partial r}|_{r=r_{+}}dr_{+}+\frac{\partial f\left(r\right)}{\partial M}|_{r=r_{+}}dM+\frac{\partial f\left(r\right)}{\partial Q}|_{r=r_{+}}dQ+\frac{\partial f\left(r\right)}{\partial l}|_{r=r_{+}}dl+\frac{\partial f\left(r\right)}{\partial\alpha}|_{r=r_{+}}d\alpha=0\text{.}\label{eq:deltaES}
\end{equation}
Putting eqns. (\ref{eq:fdrh}) into eqn. (\ref{eq:deltaES}), we can
easily get the first law of thermodynamics of Gauss-Bonnet-Maxwell
black hole in the extended phase space,
\begin{equation}
dM=\Phi dQ+TdS+VdP+\mathcal{A}d\alpha,
\end{equation}
where $dl$ is replaced with $dP$ through eqn.(\ref{eq:P}). Combining
eqn. (\ref{eq:EQES}) and the first law of thermodynamics (\ref{eq:deltaES}),
we obtain that
\begin{equation}
\left\vert P^{r}\left(r_{+}\right)\right\vert =TdS-PdV+\mathcal{A}d\alpha.\label{eq:PrES}
\end{equation}
Using eqn.(\ref{eq:P}), eqn.(\ref{eq:V}), (\ref{eq:conjugate of alpha})
and eqn.(\ref{eq:entropy}) back into (\ref{eq:PrES}), we can get
the change of the black hole entropy:
\begin{equation}
dS=\frac{\left(1+\frac{2\alpha}{r_{+}^{2}}\right)\left\vert P^{r}\left(r_{+}\right)\right\vert -\left\{ \frac{1}{2r_{+}}\left(1+\frac{2\alpha}{r_{+}^{2}}\right)+\left[\left(1+\frac{2\alpha}{r_{+}^{2}}\right)T-\frac{3r_{+}}{4\pi l^{2}}\right]2\pi\left(1-2\ln\frac{r_{+}}{\sqrt{\alpha}}\right)\right\} d\alpha}{\left(1+\frac{2\alpha}{r_{+}^{2}}\right)T-\frac{3r_{+}}{4\pi l^{2}}}.\label{eq:dsES}
\end{equation}
Based on eqn. (\ref{eq:HT}), for large enough $T$, the denominator
in eqn. (\ref{eq:dsES}) becomes
\begin{equation}
\left(1+\frac{2\alpha}{r_{+}^{2}}\right)T-\frac{3r_{+}}{4\pi l^{2}}>0.
\end{equation}
Otherwise for small enough $T$, the denominator is negative. Since
$d\alpha$ is arbitrary, the sign of the numerator in eqn. (\ref{eq:dsES})
is indefinite. In the extended phase space, the entropy can increases
or decrease depending the values of $d\alpha$. Considering the ``restricted''
extended phase space with $d\alpha=0$, the change of the black hole
entropy becomes
\begin{equation}
dS=\frac{4\pi r_{+}^{3}\left(1+\frac{2\alpha}{r_{+}^{2}}\right)\left\vert P^{r}\left(r_{+}\right)\right\vert }{\left(1+\frac{2\alpha}{r_{+}^{2}}\right)T-\frac{3r_{+}}{4\pi l^{2}}}\text{,}
\end{equation}
which shows that the second law of thermodynamics is not satisfied
for the extremal or near-extremal black hole. When the black hole
is far enough from extremality, the second law is satisfied in the
``restricted'' extended phase space.

\section{Weak Cosmic Censorship Conjecture}

\label{Sec:WCCC}

In this section, we will check the validity of the WCCC when a charge
particle enters into the black hole horizon. We assume that the initial
black hole is extremal or near extremal before absorbing a charged
particle. Since the test particle has a very small energy and charge
compared to the black hole, to become a naked singularity requires
the black hole starting to close the extremality. So here we assume
that the initial Gauss-Bonnet-Maxwell black hole is near extremal,
which possesses two horizons. Between these two horizons, there exists
one and only one minimum point at $r=r_{\min}$ for $f\left(r\right)$.
Moreover, the minimum value of $f\left(r\right)$ is not greater than
zero,
\begin{equation}
\delta\equiv f\left(r_{\min}\right)\leq0\text{,}
\end{equation}
where $\delta=0$ corresponds to the extremal black hole. After the
black hole absorbing a charged particle, the minimum point moves to
$r_{\min}+dr_{\min}$. For the final black hole solution, if the minimum
value of $f\left(r_{\min}+dr_{\min}\right)$ is still not greater
than zero, there is an event horizon. Otherwise, the final black hole
solution is over the extremal limit, and the WCCC is violated. Again,
we first present some useful formulas:
\begin{eqnarray}
\frac{\partial f\left(r\right)}{\partial r}|_{r=r_{\min}} & = & 0\text{, }\nonumber \\
\frac{\partial f\left(r\right)}{\partial M}|_{r=r_{\min}} & = & -\frac{2}{r_{\min}+\frac{2\alpha}{r_{\min}}\left(1-\delta\right)}\text{,}\nonumber \\
\frac{\partial f\left(r\right)}{\partial l}|_{r=r_{\min}} & = & -\frac{2r_{\min}^{2}}{l^{3}}\frac{1}{1+\frac{2\alpha}{r_{\min}^{2}}\left(1-\delta\right)}\text{,}\label{eq:fdrmin}\\
\frac{\partial f\left(r\right)}{\partial Q}|_{r=r_{\min}} & = & \frac{2\left(\Phi+A_{t}\left(r_{+}\right)-A_{t}\left(r_{\min}\right)\right)}{r_{\min}+\frac{2\alpha}{r_{\min}}\left(1-\delta\right)}\text{,}\nonumber \\
\frac{\partial f}{\partial\alpha}|_{r=r_{\min}} & = & \frac{\left(1-f\left(r_{\min}\right)\right)^{2}}{r_{\min}^{2}+2\alpha\left(1-\delta\right)}.\nonumber
\end{eqnarray}

\subsection{Normal Phase Space}

In the normal phase space, the charged particle with the energy $E$
and charge $q$ enters into the black hole, which makes the black
hole shift from the initial state $\left(M,Q\right)$ to the final
state $\left(M+dM,Q+dQ\right)$, where $dM$ and $dQ$ are given in
eqn. $\eqref{eq:dMdQNS}$. Moreover, the minimum value of $f\left(r\right)$
moves form $f\left(r_{\min}\right)$ to $f\left(r_{\min}+dr_{\min}\right)$,
where the final state $f\left(r_{\min}+dr_{\min}\right)$ can be rewritten
in terms of the initial state $\delta$:
\begin{eqnarray}
 &  & f\left(r_{\min}+dr_{\min};M+dM,Q+dQ\right)\nonumber \\
 &  & =\delta+\frac{\partial f}{\partial r}|_{r=r_{\min}}dr_{\min}+\frac{\partial f}{\partial M}|_{r=r_{\min}}dM+\frac{\partial f}{\partial Q}|_{r=r_{\min}}dQ\nonumber \\
 &  & =\delta-\frac{2\left\vert P^{r}\left(r_{+}\right)\right\vert }{r_{\min}+\frac{2\alpha}{r_{\min}}\left(1-\delta\right)}+\frac{2q\left[A_{t}\left(r_{+}\right)-A_{t}\left(r_{\min}\right)\right]}{r_{\min}+\frac{2\alpha}{r_{\min}}\left(1-\delta\right)}.\label{eq:fminNS}
\end{eqnarray}
Considering the extremal black hole, which implies $r_{\min}=r_{+}$
and $\delta=0$. Therefore, the minimum value of final state metric
function $f\left(r\right)$ becomes
\begin{equation}
f\left(r_{\min}+dr_{\min}\right)=-\frac{2\left\vert P^{r}\left(r_{+}\right)\right\vert }{r_{\min}+\frac{2\alpha}{r_{\min}}}<0.
\end{equation}
That is, the extremal black hole becomes non-extremal, after absorbing
a charged particle. Furthermore, if the initial black hole is near-extremal,
we define $\epsilon$ such that
\begin{equation}
r_{\min}=r_{+}\left(1-\epsilon\right)\text{,}\label{eq:infinitesimal quantity}
\end{equation}
where $0<\epsilon\ll1$. So $\delta$ is suppressed by $\epsilon$
in the near-extremal limit. Moreover, based on (\ref{eq:vector potential}),
the second term in the third line of eqn. $\eqref{eq:fminNS}$ can
be rewritten as
\begin{equation}
\frac{2q\left[A_{t}\left(r_{+}\right)-A_{t}\left(r_{\min}\right)\right]}{r_{\min}+\frac{2\alpha}{r_{\min}}\left(1-\delta\right)}=\frac{2Qq\epsilon}{r_{+}^{2}\left(1-\epsilon\right)^{2}+2\alpha\left(1-\delta\right)}.
\end{equation}
Therefore, in the near-extremal black hole, considering the test particle
limit, the third term of eqn. $\eqref{eq:fminNS}$ can be neglected.
Then eqn. $\eqref{eq:fminNS}$ becomes
\begin{equation}
f\left(r_{\min}+dr_{\min}\right)=\delta-\frac{2\left\vert P^{r}\left(r_{+}\right)\right\vert }{r_{\min}+\frac{2\alpha}{r_{\min}}\left(1-\delta\right)}<0\text{,}
\end{equation}
which means that the near-extremal black hole is still non-extremal
after the absorption. As a result, in the normal phase space, the
WCCC is satisfied for the extremal and near-extremal Gauss-Bonnet-Maxwell
black holes under the charge particle absorption.

\subsection{Extended Phase Space}

In this case, absorbing a charged particle makes the parameters of
the black hole changing from $\left(M,Q,l,\alpha\right)$ to $\left(M+dM,Q+dQ,l+dl,\alpha+d\alpha\right)$,
and $r_{\min}$ moving to $r_{\min}+dr_{\min}$. For the final state
at $r=r_{\min}+dr_{\min}$, the minimum value of $f\left(r\right)$
is
\begin{eqnarray}
 &  & f\left(r_{\min}+dr_{\min},M+dM,Q+dQ,\alpha+d\alpha\right)\nonumber \\
 &  & =\delta+\frac{\partial f}{\partial M}|_{r=r_{\min}}dM+\frac{\partial f}{\partial Q}|_{r=r_{\min}}dQ+\frac{\partial f}{\partial l}|_{r=r_{\min}}dl+\frac{\partial f}{\partial\alpha}|_{r=r_{\min}}d\alpha\nonumber \\
 &  & =\delta+\left(-TdS+q\left[A_{t}\left(r_{+}\right)-A_{t}\left(r_{\min}\right)\right]\right)\frac{2}{r_{\min}+\frac{2\alpha}{r_{\min}}\left(1-\delta\right)}\nonumber \\
 &  & \quad+\frac{r_{\min}^{3}}{l^{3}}\frac{2}{r_{\min}+\frac{2\alpha}{r_{\min}}\left(1-\delta\right)}\left[\frac{r_{+}^{3}}{r_{\min}^{3}}-1\right]dl\nonumber \\
 &  & \quad+\frac{1}{r_{\min}^{2}+2\alpha\left(1-\delta\right)}\left[\left(1-\delta\right)^{2}-\left(\frac{r_{\min}}{r_{+}}+4\pi Tr_{\min}\left(1-2\ln\frac{r_{+}}{\sqrt{\alpha}}\right)\right)\right]d\alpha,\label{eq:fminES}
\end{eqnarray}
where we use the eqn. $\eqref{eq:fdrmin}$ for the derivatives of
$f\left(r\right)$. If the initial black hole is extremal, we have
$r_{\min}=r_{+}$, $T=0$ and $\delta=0$. So we find that the minimum
value of $f\left(r\right)$ of the final black hole becomes
\begin{equation}
f\left(r_{\min}+dr_{\min}\right)=0,
\end{equation}
which means that the extremal black hole is still extremal after absorbing
the test particle. On the other hand, considering the near extremal
black hole, we define an infinitesimal quantity $\epsilon$ as (\ref{eq:infinitesimal quantity}).
And putting eqn. (\ref{eq:dsES}) into (\ref{eq:fminES}), we can
get
\begin{eqnarray}
 &  & f\left(r_{\min}+dr_{\min}\right)\nonumber \\
 &  & =\delta-\frac{2}{r_{+}\left(1-\epsilon\right)+\frac{2\alpha}{r_{+}\left(1-\epsilon\right)}\left(1-\delta\right)}T\frac{\left(1+\frac{2\alpha}{r_{+}^{2}}\right)\left\vert P^{r}\left(r_{+}\right)\right\vert }{\left(1+\frac{2\alpha}{r_{+}^{2}}\right)T-\frac{3r_{+}}{4\pi l^{2}}}\nonumber \\
 &  & \quad+\frac{1}{r_{+}^{2}\left(1-\epsilon\right)^{2}+2\alpha\left(1-\delta\right)}\left[\frac{\frac{3r_{+}}{4\pi l^{2}}}{\left(1+\frac{2\alpha}{r_{+}^{2}}\right)T-\frac{3r_{+}}{4\pi l^{2}}}+\left(1-\delta\right)^{2}\right]d\alpha\nonumber \\
 &  & \quad+\frac{\epsilon}{r_{+}^{2}\left(1-\epsilon\right)^{2}+2\alpha\left(1-\delta\right)}\left[2Qq+2\left(1-\epsilon\right)\left(\epsilon^{2}-3\epsilon+3\right)\frac{r_{+}^{4}}{l^{3}}dl-\frac{\frac{3r_{+}}{4\pi l^{2}}}{\left(1+\frac{2\alpha}{r_{+}^{2}}\right)T-\frac{3r_{+}}{4\pi l^{2}}}\epsilon d\alpha\right].\label{eq:final}
\end{eqnarray}
Since the quantity $\epsilon$ is infinitesimal, the term in the fourth
line of eqn. (\ref{eq:final}) can be neglected. However, as we discussed
in the final part of section \ref{Sec:TCPA}, in the near extremal
black hole, the black hole's temperature is small enough to make sure
the denominator in the second term of the second line of eqn. (\ref{eq:final})
is neglected. Therefore, this term is positive. Moreover, $d\alpha$
is arbitrary, hence the sign of the third line in eqn. (\ref{eq:final})
is indefinite. In general, the test particle can overcharge the near-extremal
Gauss-Bonnet black hole in 4-dimensions, which invalidates the WCCC.

\section{Conclusion}

\label{Sec:Con}

In this paper, we reviewed the solutions of the 4-dimensional Gauss-Bonnet-Maxwell black holes at first. Then, we obtained the thermodynamic quantities of the black hole and examined the first and second laws of thermodynamics by throwing a charge practice into the black hole. Finally, the WCCC for Gauss-Bonnet
black hole coupled to Maxwell theory in the normal phase space and
extended phase space were checked. Our results are summarized as follows
(\ref{tab:1}):

\begin{table}[H]
\centering%
\begin{tabular}{|p{0.6in}|p{2.6in}|p{2.6in}|}
\hline
 & Normal Phase Space & Extended Phase Space\tabularnewline
\hline
1st Law & Satisfied. & Satisfied.\tabularnewline
\hline
2nd Law & Satisfied. & Indefinite. If $d\alpha=0$ are imposed, the 2nd law is violated for
the extremal and near-extremal black holes.\tabularnewline
\hline
WCCC & Satisfied for the extremal and near-extremal black holes. After the
charge particle absorption, the extremal black hole becomes non-extremal. & Satisfied for the extremal black holes. After the charge particle
absorption, the extremal black hole stays extremal. Violated for near-extremal
black holes.\tabularnewline
\hline
\end{tabular}$\ \ $\caption{{\small{}Results for the first and second laws of thermodynamics and
the WCCC, which were tested for a }Gauss-Bonnet-Maxwell black hole{\small{}
by absorbing a charged particle in the test particle limit.}}
\label{tab:1}
\end{table}

As shown in table (\ref{tab:1}), after absorbing a charged particle, the first law of thermodynamics of the 4-dimensional Gauss-Bonnet-Maxwell black hole is still satisfied both in the normal phase space and in the extended phase space. However, the second law of thermodynamics are different in this two cases. In the normal phase space, the second law of thermodynamics is still satisfied. But the second law is indefinite in the extended phase space. More specifically, if we assume the Gauss-Bonnet parameter dose not change after the black hole absorbs a charged particle, the second law is violated for the extremal and near-extremal black holes. Furthermore, the WCCC is considered in the normal phase space. When a charged particle enters into a near-extremal black hole, the WCCC is still valid. Meanwhile, for the extremal black hole, the WCCC is violated and the black hole becomes non-extremal. On the other hand, the black hole is considered in the extended phase space. If a near-extremal black holes absorbing a charged particle, the WCCC is also still valid. However, for the extremal black hole, the WCCC is violated.

In the near future, it is reasonable to study the extended phase space thermodynamics for 4-dimensional Gauss-Bonnet-Maxwell black holes in a cavity. It motivates us to further discuss the deep relations between thermodynamic properties of novel 4-dimensional black holes and their boundary conditions. Then, based on the works of \cite{Wang:2020osg}, it is natural to discuss the validity of thermodynamic laws and the WCCC for 4-dimensional Gauss-Bonnet-Maxwell black holes in a cavity.
\vspace{5mm}

\noindent \textbf{Note:} The authors considered the same problem only
in the normal phase space which was appeared on arXiv one day ago
(April 19, 2020) \cite{Yang:2020czk}.

\vspace{5mm}

\noindent {\bf Acknowledgements}
We are grateful to Benrong Mu, Jun Tao, Peng Wang, Houwen Wu and Haitang Yang for useful discussions and valuable comments. This work is supported in part by the NSFC (Grant No. 11875196, 11375121, 11005016 and 11947225).

\end{document}